\documentclass[prx,aps,twocolumn,footinbib,superscriptaddress,floatfix,longbibliography]{revtex4-1}

\usepackage{CJK}
\usepackage[colorlinks,bookmarks=false,citecolor=black,linkcolor=black,urlcolor=black]{hyperref}
\usepackage{color} 
\usepackage{graphicx}
\usepackage{float}
\usepackage{multirow}
\usepackage{amsmath}
\usepackage{bm}
\usepackage{amsfonts}
\usepackage{braket}
\usepackage{subfigure}
\usepackage{cleveref}
\usepackage{comment}
\usepackage{dcolumn}
\usepackage{amssymb}
\usepackage{microtype}
\usepackage{xfrac}
\usepackage{array}
\usepackage{siunitx}
\usepackage[dvipsnames]{xcolor}

\newcommand{\edit}[1]{\textcolor{black}{#1}}
\newcommand{\angstrom}{\text{\normalfont\AA}}
\newcolumntype{P}[1]{>{\centering\arraybackslash}p{#1}}

\begin{document}

\title{Spin-Qubit Noise Spectroscopy of Magnetic Berezinskii-Kosterlitz-Thouless Physics}
\author{Mark Potts}
\thanks{potts@pks.mpg.de}
\affiliation{Max Planck Institute for the Physics of Complex Systems, N\"{o}thnitzer Str. 38, Dresden 01187, Germany}
\author{Shu Zhang}
\thanks{shu.zhang@oist.jp}
\affiliation{Collective Dynamics and Quantum Transport Unit, Okinawa Institute of Science and Technology Graduate University, 1919-1 Tancha, Onna-son 904-0495, Japan}
\affiliation{Max Planck Institute for the Physics of Complex Systems, N\"{o}thnitzer Str. 38, Dresden 01187, Germany}

\begin{abstract}
\textbf{Abstract:} We propose using spin-qubit noise magnetometry to probe dynamical signatures of magnetic Berezinskii-Kosterlitz-Thouless (BKT) physics. For a nitrogen-vacancy (NV) center coupled to two-dimensional XY magnets, we predict distinctive features in the magnetic noise spectral density in the sub-MHz to GHz frequency range. In the quasi-long-range ordered phase, the spectrum exhibits a temperature-dependent power law characteristic of algebraic spin correlations. Above the transition, the noise reflects the proliferation of free vortices and enables quantitative extraction of the vortex conductivity, a key parameter of vortex transport. These results highlight NV as a powerful spectroscopic method to resolve magnetic dynamics in the mesoscopic and low-frequency regimes and to probe exotic magnetic phase transitions.
\medskip

\end{abstract}

\maketitle

\begin{figure*}
    \centering
    \includegraphics[width = 0.95\linewidth]{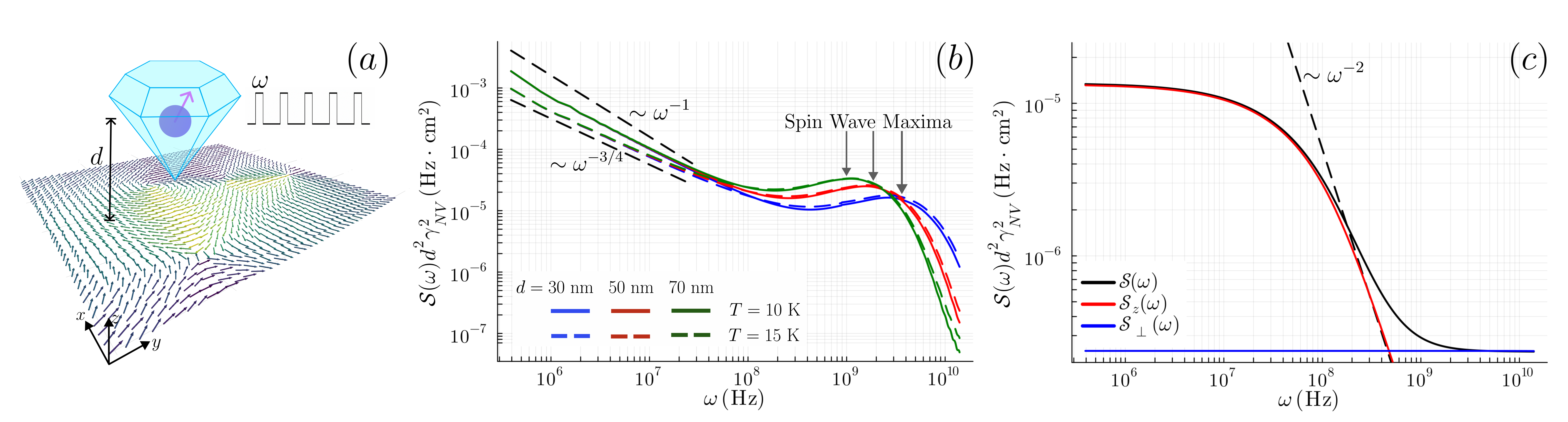}
    \caption{(a) An NV center is placed at a distance $d$ above an XY magnet and probes the magnetic noise at frequency $\omega$. (b) The noise spectral density $\mathcal{S}(\omega)$ below the BKT critical temperature $T_c=15.54$~K shows power-law behavior at low frequencies with a temperature-dependent exponent, characteristic of the algebraic spin correlations in the BKT phase. $\mathcal{S}(\omega)d^2 \gamma_\text{NV}^2$ collapses for different values of $d$, where $\gamma_\text{NV}$ is the gyromagnetic ratio of the NV electron spin. The spin wave maxima are observable at $\omega \sim c/d$, where $c$ is the renormalized bulk spin-wave velocity. (c) The noise in the disordered phase exhibits a distinctively different frequency dependence. See Eq.~(\ref{eq:rate-low-freq}). Contributions from the in-plane ($\mathcal{S}_{\perp}$) and the out-of-plane spin components ($\mathcal{S}_{z}$) alongside the full noise are plotted at $T=27.5$~K and $d=50$~nm. We have used realistic material parameters (see main text), solved the renormalization group equations, and performed integrations numerically to produce the results in (b-c).}
    \label{fig:Main_resutls_figure}
\end{figure*}

Topology plays a central role in modern condensed matter physics, particularly in identifying exotic phase transitions and unconventional forms of order. A hallmark example is the topological phase transition in two-dimensional XY systems formulated in the seminal works of Berezinskii, Kosterlitz, and Thouless~\cite{Berezinskii71,Berezinskii72,Kosterlitz73,Kosterlitz74}. A transition from quasi-long-range order to disorder is driven by the unbinding of pairs of topological defects, instead of a conventional Landau paradigm transition associated with the breaking of a continuous symmetry, which is forbidden in two dimensions at finite temperature by the Mermin-Wagner theorem~\cite{Mermin-Wagner}. This BKT transition has since been observed in various physical systems, including thin superconducting~\cite{Beasley-1979,Hebard-1980,Epstein-1981} and superfluid~\cite{Rudnick-1978,Bishop-1978} films, planar arrays of superconductor junctions~\cite{Resnick-1981}, and two-dimensional Bose gases~\cite{hadzibabic-2006,kruger-2007,clade-2009}.
However, the experimental study of magnetic BKT physics~\cite{ding-1992-bkt,Majlis-1992-dimensionalcrossover,Majlis-1993-dimensionalcrossover,Suh-1995-SCOC,Heinrich-2003-bkt,Cuccoli-2003-bkt,Carretta-2009-bkt,tutsch-2014-bkt,Kumar-2019-bkt,Hu-2020,Ashoka-2020-bkt,Caci-2021-bkt,Klyushina-2021-bkt,Opherden-2023,Zhang-Li-2024-bkt,Nakagawa-2025-bkt,Troncoso-2020,kim-2021,Flebus-2021,Seifert22,doi:10.7566/JPSJ.90.014702} has been hindered by the lack of ideal candidate materials and suitable experimental methods. 

Conventional long-range magnetic order tends to form in layered materials, even with weak inter-layer coupling, and is also precipitated by magnetic anisotropies in the XY plane.
Recent advances in the fabrication of two-dimensional van der Waals magnetic materials~\cite{li2019vdw,kurebayashi2022magnetism,park20252d} have provided promising candidates, including CrCl$_3$~\cite{zhang2015CrCl3,mcguire2015CrCl3,huang2017CrCl3,mcguire2017CrCl3} and NiPS$_3$~\cite{joy-1992-MPS3,Wildes-2015-NiPS3,kim2019NiPS3,Hu-2023-NiPS3}, that can be produced as monolayers. 
These materials possess a hexagonal magnetic planar anisotropy, which is irrelevant in the Kosterlitz-Thouless phase ~\cite{jose-1977-prb}, suggesting that the magnetic BKT transition may survive.

Noise magnetometry utilizing single-spin qubits, such as nitrogen-vacancy (NV) centers in a diamond, stands out as robust quantum sensor of local magnetic fields, suited to probe dynamics and transport in condensed matter systems~\cite{rondin-2014-nv,Casola-2018-review,xu-2023-nv-review,rovny-2024-review}. The coupling of magnetic field noise to the \edit{NV centers} drives both relaxation and dephasing processes, and measurement of the rates of these dynamics allows the extraction of frequency spectra of local magnetic noise.
Operating over nm to $\mu$m length scales and covering a broad frequency window from kHz to GHz, this approach offers complementary access to magnetic dynamics at meso- and nanoscales|bridging the gap between neutron scattering, optical probes, and transport techniques. It has been proposed and applied to study mesoscopic charge and spin transport~\cite{kolkowitz-2015-nv,ariyaratne-2018-nv,Flebus2018,Rodriguez-2018,Wang2020-nv,fang-2022-prb,zhang-2022-prb,Rodriguez-2022,xue24,PhysRevB.95.155107}, dynamics of topological defects~\cite{dussaux-2016-nv,Flebus2018,Flebus-2018-dm,Jenkins-2019-nv,Juraschek-2019-dm,mclaughlin-2022-mbt,rable-2023-dm,mclaughlin-2023-dm,Schlussel-2018-sc}, and dynamical phenomena in exotic phases and phase transitions~\cite{Chatterjee-2019,konig-2020-sc,Chatterjee-2022-sc,Machado-2023-prl,Dolgirev-2024-wigner,curtis2024probing,liu2025,PhysRevResearch.6.013043,PhysRevB.98.195433,de2025nanoscaledefectsprobestime}. 

In this work, we propose leveraging the capabilities of NV noise magnetometry to investigate the dynamical features of magnetic BKT physics. We calculate the magnetic noise spectral density in the MHz$\sim$GHz regime, which can be measured by an NV center in proximity to an XY magnet. In the BKT phase, we find a characteristic power-law spectrum with a temperature-dependent exponent, a distinctive hallmark of the algebraic spin correlations intrinsic to the quasi-long-range order. We also predict the functional form of the noise spectrum in the high-temperature disordered phase resulting from spin waves overdamped by free vortices. This offers a noninvasive method to extract the vortex conductivity and quantify vortex dynamics, which can be generally applied to other magnetic systems. The temperature variations in the spectrum clearly capture the proliferation and dynamics of vortices that drive the topological phase transition. Our results highlight the direct access of spin-qubit noise magnetometry to magnetic dynamics in the mesoscopic and low-frequency regimes and promote the use of NV centers as a spectroscopic tool in condensed matter and material studies. 

\textit{Main Results.}|The NV relaxation and decoherence times ($T_1$ and $T_2$) measure the magnetic noise, i.e. temporal fluctuations of the magnetic field, 
at the position of the NV center that is produced by the spin dynamics within the material system under study. Our main objective is thus to compute the noise spectral density $\mathcal{S} (\omega)$ from the spin correlations functions of an XY magnet.
For a two-dimensional system with axial symmetry, we can separate out the contributions from out-of-plane and in-plane spin components:
$\mathcal{S} (\omega) = \mathcal{S}_z (\omega) + \mathcal{S}_\perp (\omega)$,
where for $\mu = z, \perp$,
\begin{equation}
    \mathcal{S}_{\mu} (\omega) = \gamma^2 f(\theta_{NV}) \int \mathrm{d} k \, k^3 e^{-2kd}\mathcal{C}_{\mu} (\omega, k).
    \label{eq:noise-spetral-density}
\end{equation}
Here, $\mathcal{C}_z (\omega, k)$ is the Fourier transform of $\mathcal{C}_z (t,r) \equiv \langle S_z (t, r) S_z (0,0)\rangle$, and $\mathcal{C}_\perp (\omega, k)$ that of $\mathcal{C}_\perp (t,r) \equiv \langle \sum_{i = x,y} S_i (t, r) S_i (0,0)\rangle/2$. $\gamma$ is the gyromagnetic ratio of the magnet, and $k^3 e^{-2kd}$ is the form factor associated with the stray field. $d$ is the vertical distance from the NV center to the plane of the XY magnet, which defines the length scale and corresponds to a wavevector $k \sim 1/d$ of spin fluctuations the probe is most sensitive to. The geometric factor $f(\theta_{NV})$ depends on the orientation of the NV spin, and whether we measure the stray field noise transverse or longitudinal to the NV axis, these being accessed by $T_1$ and $T_2$ measurements respectively. 

The main results of this work are the distinctive features in the magnetic noise spectrum for the quasi-long-range ordered BKT phase and the high-temperature vortex plasma, as summarized in Fig.~\ref{fig:Main_resutls_figure}. We plot $\gamma_{NV}^2 \mathcal{S} (\omega)$, in units of Hz, to facilitate direct comparison with experiments. The overall factor of $\gamma^2 f(\theta_{NV})$ in Eq.~(\ref{eq:noise-spetral-density}) will be implicit in the text from now on. 
Focusing on the the low-frequency regime $\omega \ll c/d$, where $c$ is the spin wave velocity of the XY magnet, we observe a clear change in spectral behavior above and below the BKT transition:
\begin{equation}
    \mathcal{S} (\omega) \sim \left\{
    \begin{array}{lll}
        \omega^{\eta -1}, \quad & T \! \lesssim \! T_c, \\
        1/ \left( 1+ \Omega^2 \omega^2/\omega_s^4 \right), \quad &  T \! > \! T_c.
    \end{array}
    \right.
    \label{eq:rate-low-freq}
\end{equation}
\edit{$\Omega$ is an emergent plasma frequency, and $\omega_s$ a resonant spin wave frequency that shall be defined shortly.} Below the transition temperature $T_c$, the noise spectrum exhibits a power-law behavior across several orders of magnitude in frequency, as shown in Fig.~\ref{fig:Main_resutls_figure} (b). This behavior is inherited directly from the algebraic spin correlations associated with the quasi-long-range order and serves as a clear signature of BKT physics. The exponent of this power law, $\eta-1$, is governed by the dimensionless parameter $\eta = k_B T / 2\pi J$, with $J$ the renormalized spin stiffness in the long-wavelength limit. Approaching the critical point $T_c$, $\lim_{T \rightarrow T_c^-}\eta = \eta_c=1/4$, and the low-frequency exponent drifts from $-1$ towards $-3/4$. 

Above $T_c$, spin dynamics are governed by the proliferation of free vortices. Their behavior is analogous to a plasma, with an associated emergent plasma frequency $\Omega = 2\pi\sigma/\epsilon_c$. Here $\epsilon_c$ is the bulk critical value of the \textit{emergent} dielectric constant, and $\sigma = 2\pi \nu J_0 n_f$ is the vortex ``conductivity'' [See Eq.~(\ref{eq:Plasma_current})] dependent on the free vortex density $n_f$ and the vortex mobility $\nu$. As with propagating light in a traditional plasma, the response of free vortices to spin waves above $T_c$ overdamps these modes at frequencies below $\Omega$, whilst those at higher frequencies continue to propagate.
At temperatures somewhat above $T_c$, where the free vortex density is sufficiently large that the plasma frequency is above the resonance peak \edit{$\omega_s$} of the linearly-dispersed spin wave, $\Omega \gg \omega_s \sim 5c/2d$, the low-frequency spectral features are generated by overdamped spin wave modes, as given in Eq.~\ref{eq:rate-low-freq} and plotted in Fig.~\ref{fig:Main_resutls_figure} (c). Fitting of the measured noise spectrum allows the extraction of the plasma frequency, and hence the vortex conductivity.  

\textit{Model.}| Our analysis of magnetic noise from an XY magnet begins with the following model Hamiltonian:
\begin{equation}
    H=\frac{J_0}{2S^2}\sum_{i = x,y}  \int \text{d}^2\mathbf{r} \, (\bm{\nabla}S_i)^2 + \frac{1}{2\alpha}\int\text{d}^2\mathbf{r} \ S_z^2.
    \label{eq:Hamiltonian}
\end{equation}
Here $J_0$ is the exchange stiffness, and $\alpha$ is an easy-plane anisotropy that keeps spins predominantly in the $xy$ plane. The $S_{\mu}$ are coarse-grained spin densities and $S$ is the saturated spin (angular momentum) density. For this model to be well described by the planar XY model up to a momentum scale $k_{\text{max}}$, one requires $1/\alpha \gg J_0k_{\text{max}}^2/S^2$. This can be achieved with only mild magnetic anisotropy (including the single-ion and magnetic dipolar effects) provided one is not probing scales comparable with the lattice spacing $a_0$.

The azimuthal angle $\phi$ parameterizes the in-plane spin components $S_x=S\cos\phi$ and $S_y=S\sin\phi$, and 
its dynamics are accompanied by a small tilt out of the plane $S_z=\alpha \dot{\phi}$ (following from the canonical Poisson brackets for spins). 
This XY magnet can then be mapped to electromagnetism in ($2 + 1$) dimensions~\cite{Kosterlitz73,Ambegaokar80,Cote86,Dasgupta-2020}: The electric and magnetic fields in terms of the field $\phi(t,\mathbf{r})$
\begin{equation}
    \mathbf{E}=\sqrt{2 \pi J_0} \ \bm{\nabla} \phi \times \hat{\mathbf{z}}, \; \text{and} \;
    B = \sqrt{2 \pi \alpha} \ \dot{\phi} ,
    \label{eq:fields}
\end{equation} 
satisfy a complete analog of the Maxwell equations, with charge density $\rho$ and current density $\mathbf{j}$ proportional to the vortex number density $n_f$ and current $\mathbf{j}_v$ respectively~\cite{suppl}. The equation of motion for $\phi$, corresponding to the Ampère-Maxwell law, describes linearly dispersing spin waves, or equivalently an emergent photon, with bare speed $c_0=\sqrt{J_0/\alpha}$. 

Vortex defects play an essential role in driving the BKT transition, and to model their behavior correctly, $\phi$ is split into a smooth part $\theta$ associated with spin waves, and a singular part $\psi$ associated with the vortices. This is equivalent to a Helmholtz decomposition $\mathbf{E} = \mathbf{E}_T + \mathbf{E}_L$, with the transverse component (to the wavevector $\mathbf{k}$) being  $\mathbf{E}_T \propto \! \bm{\nabla} \theta \times \hat{\mathbf{z}}$ and the longitudinal component given by $\mathbf{E}_L \propto \! \bm{\nabla} \psi \times \hat{\mathbf{z}}$.
As vortex cores behave like charges, they have an associated logarithmic interaction energy. Above a critical temperature $T_c$, the gain in entropy for unbinding vortex-antivortex pairs exceeds the unbinding energy cost. The proliferation of free vortices is then responsible for the destruction of the low temperature quasi-long-ranged order~\cite{Berezinskii71,Berezinskii72,Kosterlitz73,Kosterlitz74}.

Below $T_c$, the bound vortex pairs serve as instantaneous dipoles, giving the system a finite, scale dependent polarizability, and hence a dielectric constant $\epsilon(r)$. This is the central mechanism behind the renormalisation group analysis of Berezinskii, Kosterlitz, and Thouless ~\cite{Berezinskii71,Berezinskii72,Kosterlitz73,Kosterlitz74,Young78}. $\epsilon(r)$ renormalizes the spin stiffness and the spin wave velocity. 
The scale dependent dielectric constant can be made dynamical through consideration of the response of vortex pairs to a perturbing potential ~\cite{suppl}. $\epsilon (\omega, k)$ is calculated by modeling vortex kinetics with the assumption that drag forces \edit{originating from the Gilbert damping} dominate over the emergent Lorentz force. Whilst in this work we have neglected the spin wave broadening directly due to Gilbert damping, it can in principle be included as a small correction to the imaginary part of $\epsilon(\omega, k)$. The drifting motion of the (massless) vortices can then be described by a Langevin equation~\cite{Ambegaokar80} with an approximately constant vortex mobility $\nu$~\cite{PhysRev.140.A1197,curtis2024probing,Halperin79,Huber82}.

Above $T_c$, pairs remain bound only within a correlation length $\xi_+\sim \exp(b/\sqrt{T-T_c})$. The dynamical dielectric constant involves a bulk contribution saturated at the critical value $\epsilon_c$ from the remaining bound vortex pairs, and a contribution from the free vortex current~\cite{suppl}:
\begin{equation}
\mathbf{j}_{\text{free}}(\omega,k)=\sigma\frac{1}{1+\mathrm{i} Dk^2/\omega}\mathbf{E}_L+\sigma\mathbf{E}_T.
\label{eq:Plasma_current}
\end{equation}
Here we define $\sigma = 2\pi \nu J_0 n_f$ as the vortex conductivity, which is proportional to the density of free vortices $n_f \sim 1/\xi_+^2$ and, via the Einstein relation, to the vortex diffusion constant $D = \nu k_B T$ . 
The longitudinal electric field is screened by the diffusive vortices, giving an incompressible current in the static limit $\omega\rightarrow 0$.

\begin{figure}
    \centering
    \includegraphics[width = \linewidth]{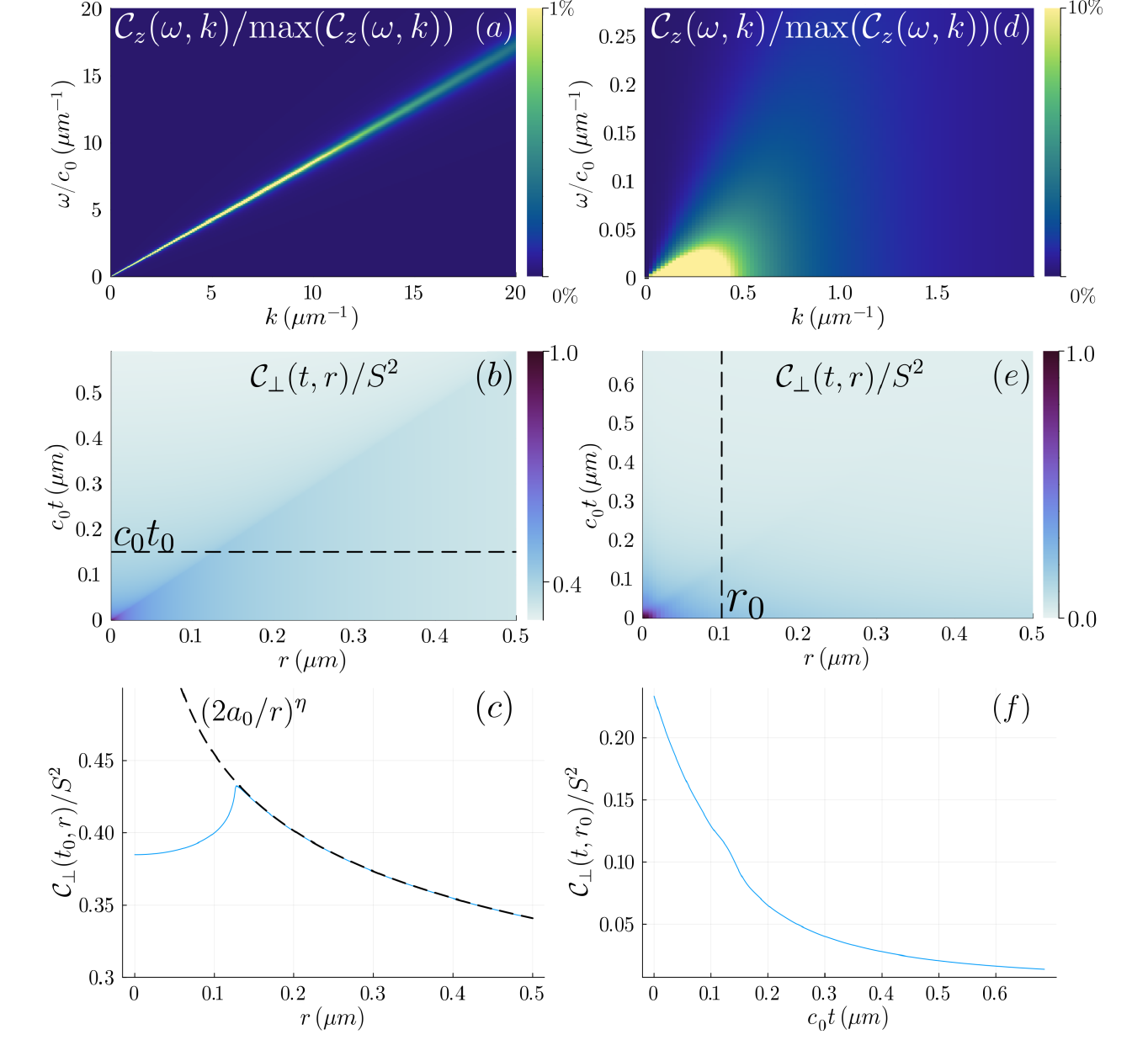}
    \caption{In-plane and out-of-plane spin correlation functions below $T_c$ (a-c) and above $T_c$ (d-f). Here, $J_0/k_B \sim 10$~K and $\mu_0 \sim 2 J_0$ are used to enhance the renormalization effects. (a) Below $T_c$, the out-of-plane spin correlations in the momentum-frequency space $\mathcal{C}_z (\omega, k)$ show a clear linearly-dispersed spin wave, which increasingly broadens at higher $k$. (b) Above $T_c$, spin wave modes below the plasma frequency are overdamped by free vortices. Panels (b) and (e) present the in-plane correlations above and below $T_c$. (b) Below $T_c$, the in-plane spin correlations in the real time and space $\mathcal{C}_\perp (t, r)$ has a line of maxima alone the $c_0 t/\sqrt{\text{Re}\, \epsilon (r)}=r$ with power-law decays on either side, as shown in the cut (c). (e) Above $T_c$, the algebraic correlations are washed away, leaving an exponential decay shown in cut (f).}
    \label{fig:Correlation_functions}
\end{figure}

\textit{Signatures of BKT phase.}|Following the BKT phenomenology presented above, we apply standard techniques from electromagnetism to obtain the spin-density correlations and the resulting magnetic field noise spectrum~\cite{suppl}. The starting point is the (retarded) Green's functions $\mathcal{G}(\omega, k)$ (isotropic in the $xy$ plane) of the vector potential of the fields~(\ref{eq:fields}), which follow directly from Fourier transforming the Maxwell equations.
In the low-temperature BKT phase with no free vortices, only the transverse response is present:
\begin{equation}
    \mathcal{G}_T (\omega, k) = \frac{2\pi \hbar}{\left[ \left(\omega^2/c_0^2 \right) \epsilon(\omega, k)\right] - k^2}.
\end{equation}

Invoking the fluctuation-dissipation theorem in the classical regime $\hbar \omega \ll k_B T$ (satisfied for a probing frequency $\sim$ MHz-GHz and temperature $\sim 10$~K), one obtains the correlation function of the transverse vector potential, and hence those of the electromagnetic fields, which can in turn be related to the correlation functions of the order parameter field $\phi$ and of $S_z$. Using $S_z = \sqrt{\alpha/2\pi} B$, one obtains  $\mathcal{C}_z (\omega, k) = \left( k_BT J_0 k^2/\hbar \pi c_0^2 \omega  \right) \text{Im} \mathcal{G}_T (\omega, k)$. 
As shown in Fig.~\ref{fig:Correlation_functions}.(a), for $T<T_c$, $\mathcal{C}_z (\omega, k)$ peaks along the spin wave dispersion with renormalized velocity $c(\omega)=c_0/\sqrt{\text{Re}\,\epsilon(\omega,k \rightarrow 0)}$ (we assume the spin wave wavelength is much larger than the distance $\sim \sqrt{D/\omega}$ that vortices can diffuse within the wave period) and broadens as $\sim k^{\pi J/{k_BT}-1}$~\cite{Ambegaokar80}. This renormalization is relatively weak if a realistic value for the bare vortex chemical potential $\mu_0$ is taken to be several times of $J_0$ \cite{Kosterlitz73}. Since this intrinsic broadening is much narrower than the width of the wavevector form factor of NV centers, we can approximate $\mathcal{C}_z (\omega, k)$ by a line of delta functions at $k = \omega/c(\omega)$,  which gives a clean form of the noise spectral density:
\begin{equation}
    \mathcal{S}_z(\omega) \sim  \pi k_BT \frac{ J_0}{c_0^2}  
    \frac{ \omega ^3}{c^4(\omega)}e^{-\frac{2 \omega d}{c(\omega)}}, \label{eq:Gamma_z_low_T}
\end{equation}
with a maximum at $\omega \sim 3c(\omega)/2d$, visible in Fig.~\ref{fig:Main_resutls_figure} (b).

The in-plane spin correlations are computed via  $\mathcal{C}_\perp (t,r) = S^2 \exp \{ -\langle [\phi(t,r) - \phi(0,0)]^2 \rangle/2 \}$. Given that $\text{Im}\,\epsilon (\omega, k)$ is small~\cite{suppl}, we have the following analytical expression for the long-wavelength (much larger than lattice spacing $a_0$) scaling behavior:
\begin{align}
    &\mathcal{C}_{\perp}(t,r)
     \approx
    S^2\left(\frac{2a_0}{r}\right)^{\eta}\Phi\left(\frac{ct}{r}\right),  \nonumber \\ 
   & \text{where} \quad \Phi(u) = \begin{cases}
    1,\ \ 0<u<1, \\
    [u+(u^2-1)^{1/2}]^{-\eta}, \ \ u>1.
\end{cases}
    \label{eq:In-plane-correlations}
\end{align}
Here, $c = c_0/\sqrt{\epsilon_\infty}$, where $\epsilon_{\infty}$ is the renormalized bulk dielectric constant. The scaling exponent is given by the dimensionless temperature $\eta=k_BT/2\pi J$. $\mathcal{C}_{\perp}(t,r)$ peaks at $r = c \, t$ and shows algebraic decay on both sides, characteristic of the quasi-long-ranged order of the low temperature BKT phase [See Fig.~\ref{fig:Correlation_functions}(b, c)]. Numerical integration is performed to obtain the Fourier transform $\mathcal{C}_\perp (\omega, k)$ and the noise spectral density $\mathcal{S}_{\perp}(\omega)$, which is shown in Fig.~\ref{fig:Main_resutls_figure}(b). The low-frequency scaling behavior can be seen by an approximation taking $r \sim d$ in Eq.~(\ref{eq:In-plane-correlations}), resulting in
\begin{equation}
    \mathcal{S}_{\perp}(\omega) \sim \frac{\pi S^2}{2d^2} \int_0^{\infty}\text{d}t \cos(\omega t) \left(\frac{a_0}{ct}\right)^{\eta}.
\end{equation}
This integral over time yields a power law $\omega^{\eta-1}$. As shown in Fig.~\ref{fig:Main_resutls_figure}(b), $\mathcal{S}_{\perp}(\omega) d^2$ curves at various values of $d$ collapse well for $\omega \ll c/d$, and show a clean power law across orders of magnitude in frequency, with the temperature dependent exponent $\eta - 1 \in [-1,-3/4]$. 
This spectral dependence serves as a distinct signature for the algebraic spin correlations in the quasi-long-range-ordered BKT phase, in contrast to the magnetic noise spectra below the spin wave gap previously studied in long-range-ordered magnetic systems, where magnon diffusion typically leads to a $\omega^{-2}$ dependence~\cite{zhang-2022-prb,fang-2022-prb}, or a peak may arise at a collective magnon hydrodynamic mode from magnon interactions~\cite{Rodriguez-2022,xue24}. 
\edit{Such an exponent can also distinguish the BKT phase from the behavior of critical noisy dynamics near continuous phase transitions~\cite{Machado-2023-prl}. }
In the clean thermodynamic limit, the zero frequency noise diverges, which is regulated by the finite system size $L$, with the scaling $\mathcal{S}_{\perp}(\omega \rightarrow 0) \sim L^{1-\eta}$~\cite{suppl}.

{\it Vortex conductivity in the disordered phase.}|We next turn to the disordered phase above $T_c$, where the free vortex-current plays an essential role. The retarded Green's function of the vector potential now has both transverse and longitudinal components:
\begin{align}
    \mathcal{G}_T (\omega, k) &= \frac{2\pi \hbar}{\left[ \left(\omega^2/c_0^2 \right) \epsilon(\omega, k)\right] - k^2 + \mathrm{i} 2\pi \sigma \omega /c_0^2 }, 
    \label{eq:transverse-response-above-tc}\\
     \mathcal{G}_L (\omega, k) &= \frac{2\pi \hbar}{\left(\omega^2/c_0^2 \right) \left[  \epsilon(\omega, k)- 2\pi \sigma/ \left( \mathrm{i} \omega - Dk^2\right) \right]  }.
     \label{eq:longtitudinal-response-above-tc}
\end{align}
For temperatures modestly higher than $T_c$, a constant approximation $\epsilon (\omega, k) \approx \epsilon_c$ works well, as the correlation length quickly approaches the scale of $a_0$ as temperature increases. Below, we use the $\epsilon_c$ in the formulas for brevity, while the frequency and momentum dependence is retained in the numerical computations.
We identify a frequency scale in Eq.~(\ref{eq:transverse-response-above-tc}) that plays the role of a plasma frequency: $\Omega =  2\pi\sigma/\epsilon_c$. For $\omega<\Omega$, the dispersive transverse spin wave is predominately relaxational, $\omega_{T} \approx \mathrm{i} \left(c_0^2/2\pi \sigma \right) k^2$. At long length scales, $\Omega$ also determines the relaxation rates of the second transverse mode $\omega'_{T} \approx -\mathrm{i} \Omega$ and the longitudinal mode $\omega = -\mathrm{i} \left( D k^2 + \Omega \right)$, which is overdamped due to vortex diffusion. Consequently, the spin correlations in the low-frequency regime lose any sharp features with a broadening $\sim \Omega$. 

\begin{figure}
    \centering
    \includegraphics[width =\linewidth]{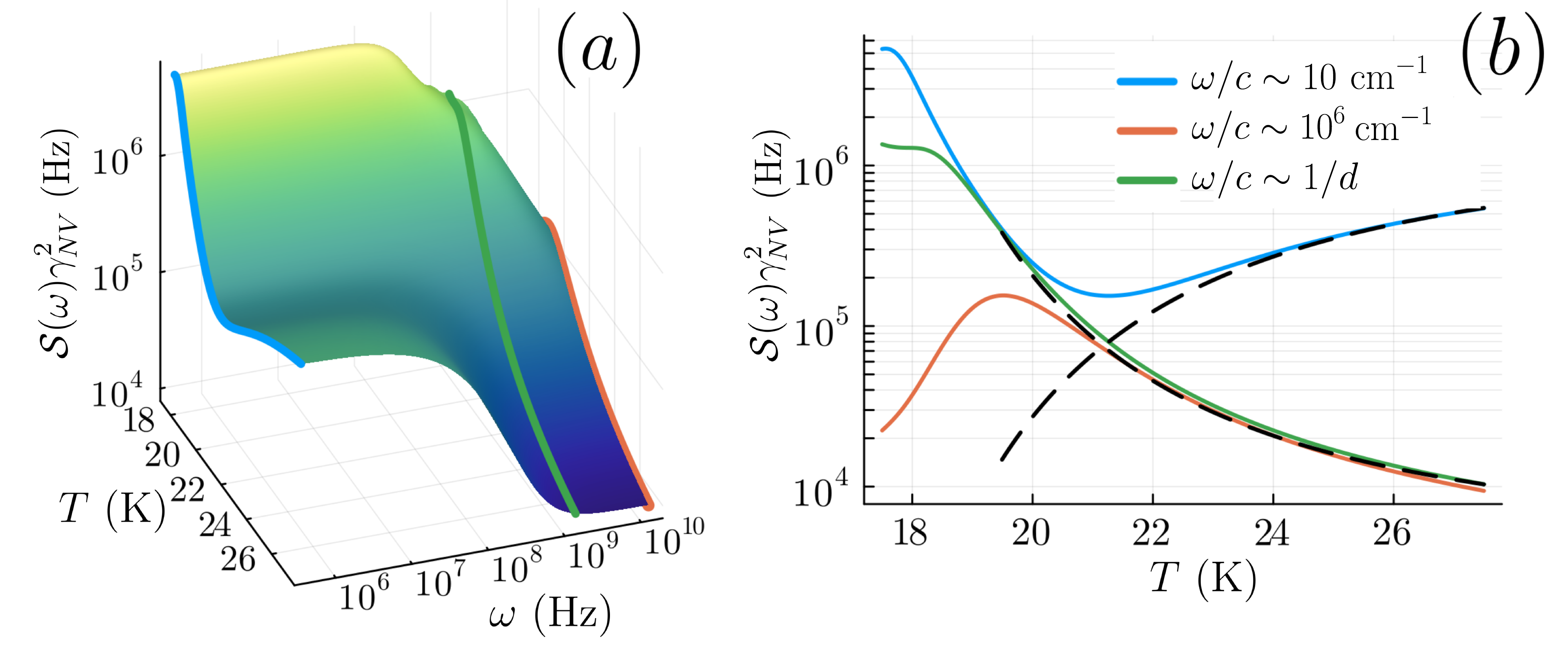}
    \caption{(a) The magnetic noise spectral density for a range of temperatures above $T_c$ and (b) selective cuts at fixed frequencies. $d=50$~nm is used. Residual spin wave maximum only visible at the low-frequency end of the green curve with $\omega \sim c/d$ and is absent at higher temperatures. Accompanying the suppression of the spin wave maximum, the spectral weight shifts into a low-frequency plateau with increasing $T$. The dashed lines in (b) show $\pm\log_{10}(D n_f)$ up to a constant shift, which describes the high-temperature behavior of the noise spectrum at low and high frequencies, respectively, inheriting the temperature dependence from the vortex diffusion constant $D$ and vortex density $n_f$.}
    \label{fig:Above_Tc_cuts}
\end{figure}

The out-of-plane spin correlations result only from the spin wave dynamics, $\mathcal{C}_z (\omega, k) = \left( k_BT J_0 k^2/\hbar \pi c_0^2 \omega  \right) \text{Im} \mathcal{G}_T (\omega, k)$, displaying overdamped modes for low frequency [Fig.~\ref{fig:Correlation_functions} (d)]. Above the plasma frequency $\omega >\Omega$,  propagating spin waves remain and contribute to the noise spectrum as peaks in the high-frequency, low-temperature regime. On the other hand, vortex diffusion completely removes the algebraic scaling structure from the in-plane spin correlations for length scales above the correlation length~\cite{Huber82,Cote86} (to calculate these for $k\lesssim1/\xi_+$, the vortex field $\psi$ can be interpreted as arising from a series of vortex multiplets, and is well defined, allowing one to compute the contribution to in-plane spin correlations from the longitudinal response of the vector potential). The in-plane spin correlations are:
\begin{equation}
    \mathcal{C}_{\perp}(t) \sim
    S^2e^{-4\pi Dn_f\ln \left(L/a_0 \right) t} \label{eq:High_T_in_plane_correlations},
\end{equation}
which decays exponentially in time [Fig.~\ref{fig:Correlation_functions} (f)] with a lifetime logarithmically diverging with the system size. Indeed, the system is disordered and uncorrelated in any macroscopic scale. In the noise spectrum, the exponential decay gives an extremely broad Lorentzian centered at zero: 
\begin{equation}
    \mathcal{S}_\perp (\omega) \sim S^2 \frac{F(a_0, d) }{d^2}\frac{W}{\omega^2 + W^2},
\end{equation}
where the half-width at half-maximum is $W = 4\pi D n_f \ln (L/a_0)$ and $F(a_0,d) \approx (3\pi/8d^2)\int_0^{\infty}\text{d}r \, r \\ \left(1-3r^2/8d^2\right) \! \left(1+r^2/4d^2\right)^{-7/2} \!\left(2a_0/r\right)^{\eta_c}$. This Lorentzian can be seen as a constant plateau for all of the relevant frequencies [Fig.~\ref{fig:Main_resutls_figure} (c)], corresponding to the high-frequency plateau in Fig.~\ref{fig:Above_Tc_cuts}, with a temperature-dependent height $\sim 1/ D n_f $. 

The low-frequency noise spectrum results from the overdamped spin modes in $\mathcal{C}_z (\omega, k)$. Taking $k\sim 5/2d$, where the momentum form factor $k^5 \exp(-2kd)$ peaks, 
\begin{equation}
    \mathcal{S}_z(\omega)  \sim 
   \frac{ 24 \pi\sigma k_BT J_0}{125 d^2 c_0^4}
    \frac{1}{1
    \!+\! \left[(\Omega/\omega_s)^2 \!-\! 2\right](\omega/\omega_s)^2
    \!+\!(\omega/\omega_s)^4} ,
    \label{eq:noise-above-Tc}
\end{equation}
where $\omega_s = 5c_0/2d\sqrt{\epsilon_c}$. We can therefore fit the measured noise spectrum to this form (\ref{eq:noise-above-Tc}) with two the fitting parameters $\Omega$ and $\omega_s$, besides an overall factor and a constant background, to extract the vortex conductivity $\sigma = (\Omega/2\pi) (5c_0/2d \omega_s)^2$. In the regime of low-frequency $\omega \ll \omega_s$ and high temperature with a sufficiently large vortex density such that $\Omega \gg \omega_s$, Eq.~(\ref{eq:noise-above-Tc}) reduces to the simpler form presented in the main results Eq.~(\ref{eq:rate-low-freq}). As shown in Fig.~\ref{fig:Main_resutls_figure} (c), the drop in the noise spectrum with increasing frequency is proximate to a $\omega^{-2}$ dependence in the window $\omega_s^2/\Omega \ll \omega \ll \omega_s$.
The low-frequency plateau, on the other hand approaches the value $ \mathcal{S}_z(\omega \rightarrow 0)$, which has a temperature dependence following $D n_f$, as shown in Fig.~\ref{fig:Above_Tc_cuts}. Noting that $\sigma = (2\pi J_0 /k_B T) Dn_f$, the temperature-dependent plateaus provide an additional reference for the $\sigma$ extracted from the spectral fitting at different temperatures.

{\it Predictions for a van der Waals ferromagnet.}|To provide quantitative predictions for an experimental NV measurement, Fig.~\ref{fig:Main_resutls_figure} and \ref{fig:Above_Tc_cuts} are plotted with material parameters relevant for a van der Waals ferromagnet. We take the bare spin stiffness $J_0/k_B \sim 10$~K, and a conservatively small easy-axis anisotropy $\hbar^2/\alpha a_0^2 k_B \sim 0.32$~K, where the lattice constant $a_0 \sim 6~{\angstrom}$. For example, the anisotropy is estimated to be $\sim 1.3$~K for NiPS$_3$ \cite{Hu-2023-NiPS3} and $\sim 6.3$~K for TmMgGaO$_4$ \cite{Hu-2020}. Here, the bare spin wave velocity is $c_0 \sim 1.4 \times 10^4$~cm/s and the BKT transition temperature is $T_c = 15.54$~K. The bare vortex chemical potential is set at $\mu_0 \sim \pi^2 J_0$~\cite{Kosterlitz73}, and the vortex density is computed by running the renormalization group flow. The vortex mobility is taken to be $\nu \sim 6.1 \times 10^{9}$~s/g~\cite{Huber82}, yielding a vortex diffusion constant $D\sim2.3\times10^{-5}$~cm$^2/$s at $T=27.5$~K.
In Fig.~\ref{fig:Main_resutls_figure} (c), we have $\omega_s\sim7.0$~GHz, $\Omega\sim54.7$~GHz, and $\sigma\sim8.8$~GHz. For an experimentally measured noise spectral curve, these parameters can be obtained fitting to Eq.~(\ref{eq:noise-above-Tc}). In the kHz-GHz frequency window, the magnetic noise signal has an order of magnitude of $10^4$-$10^7$~Hz, which is well within the experimental reach. 

\textit{Discussion.}|We have demonstrated that spin-qubit magnetic-noise spectroscopy offers a unique probe to identify the dynamical features in magnetic BKT physics. This technique directly accesses the algebraic spin correlations and spin-wave excitations of the quasi-long-ranged order of the BKT phase, and further enables quantitative measurement of the vortex conductivity in the disordered regime. Experimentally, promising platforms are monolayer van der Waals magnets with hexagonal lattice symmetry. Although these systems are expected to undergo a magnetic ordering transition that spontaneously breaks the sixfold symmetry, the BKT transition may occur at a temperature higher than this~\cite{Nelson77}, and the XY-type of dynamics can be relevant for a wider temperature range at finite wavevectors~\cite{Seifert22}. Importantly, the prominent wavevector under probe is set by the NV–sample distance, allowing access the low-frequency features while insensitive to lattice-specific details. Our theoretical framework extends naturally to antiferromagnetic systems, and the predicted behaviors for the magnetic noise remain qualitatively valid for systems with an Néel order parameter that does not enlarge the crystallographic unit cell.

\edit{A small magnetic field needs to be applied along the NV axis to tune the NV resonance frequency. Due to the sensitivity of the BKT transition is to in-plane magnetic fields that break U(1) symmetry, the preferred measurement geometry is to align the NV axis to the z direction.
Noting that the energy scale for the applied fields is of the order of GHz $\sim 0.01$K \cite{doi:10.1126/science.1192739}, much lower than typical magnetic exchange and anisotropy energies of the order THz $\sim 10$K \cite{Hu-2020,Hu-2023-NiPS3}, the BKT transition will not much affected~\cite{PhysRevB.104.064402,PhysRevB.106.214402,PhysRevLett.130.086704}.}

Finally, we contrast the magnetic-noise signatures of BKT physics in magnets with those in two-dimensional superconductors. A recent study~\cite{curtis2024probing} demonstrates that NV magnetometry can probe the dynamical dielectric function across the superconducting BKT transition, where magnetic noise is dominated by electric current fluctuations and exhibits a cusp-like feature. In the magnetic case, however, the noise arises from the dynamics of the order parameter itself, and no sharp singularity is expected at the transition. Instead, the rapid growth of the free-vortex density drives a steep increase in the plasma frequency, leading to a pronounced redistribution of spin-wave spectral weight from high to low frequencies. The spectral dependence enables direct characterization of the scaling behavior of spin correlations in the low-temperature phase and vortex transport phenomena in the high-temperature phase. These features establish NV-based measurements in magnetic systems as a powerful spectroscopic tool to quantitatively resolve dynamical properties and reveal their changes across phase transitions, complementary to the thermodynamic singularities.

\section*{Supporting Information}
(I) A review of the formulation of the XY model in terms of emergent electromagnetism in two-dimensions; (II) details of the kinetic theory used to describe the dynamics of vortices in this work; (III) details of the linear response theory used to determine spin correlation functions; (IV) a short discussion on naive finite size effects on algebraic correlations. 

\section*{Acknowledgements}
The authors thank Mengxing Ye, Jonathan Curtis, and Roderich Moessner for inspiring discussions. This work was supported in part by
the Deutsche Forschungsgemeinschaft under Grant No. SFB
1143 (Project-ID No. 247310070) and by
the Deutsche Forschungsgemeinschaft  under cluster of excellence
ct.qmat (EXC 2147, Project-ID No. 390858490). 

\section*{References}
\bibliography{Bibliography.bib}
\nocite{Ambegaokar79}
\nocite{Lifshitz80}

\newpage
\textbf{For Table of Contents Only:\\ \\ }
\begin{figure}[h!]
    \includegraphics[width = 3.25in]{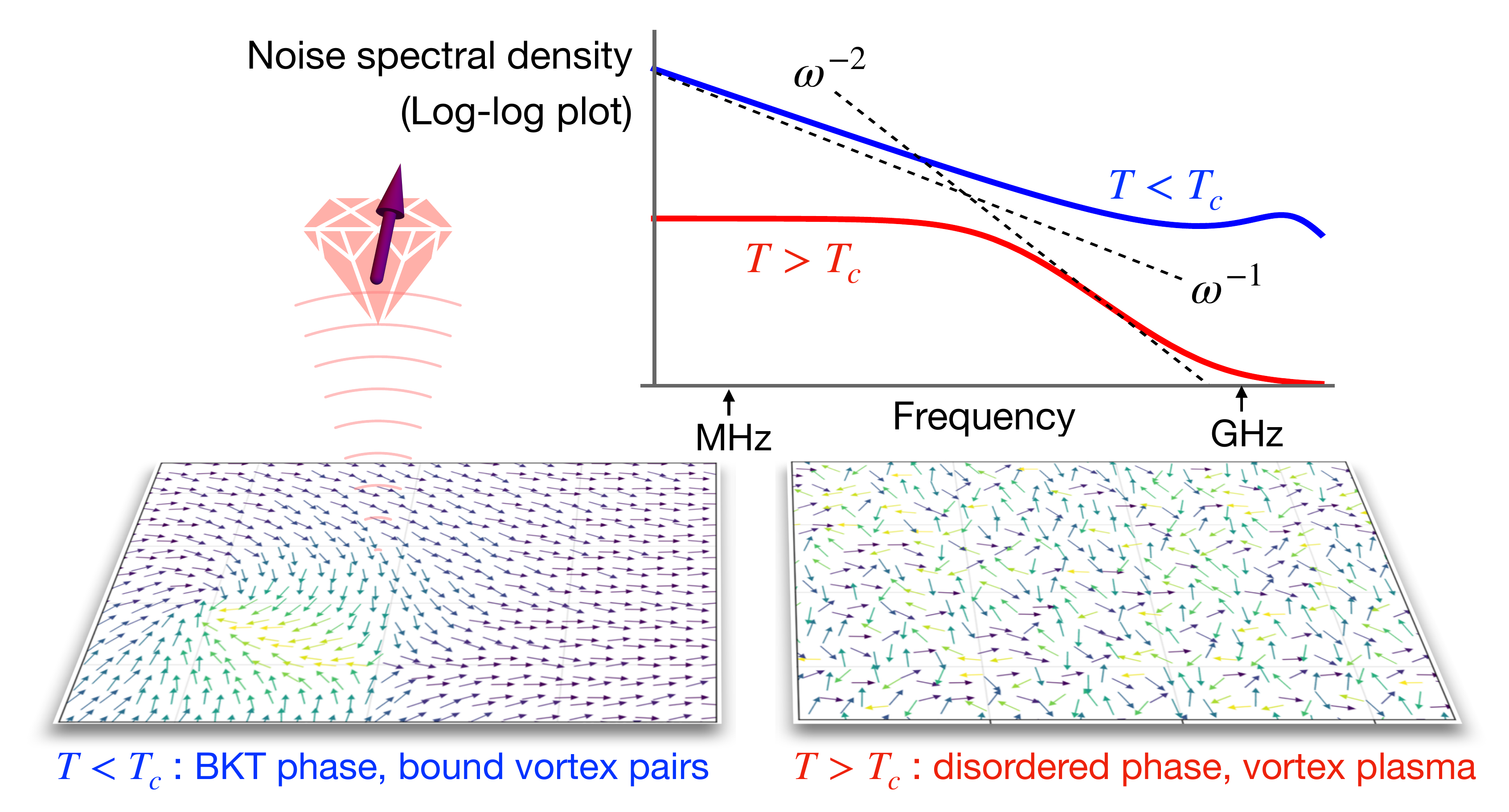}
\end{figure}

\end{document}


\title{\vspace{-2cm}Supporting Information for \\ `Spin-qubit Noise Spectroscopy of Magnetic Berezinskii-Kosterlitz-Thouless Physics' }
\title{Spin-qubit Noise Spectroscopy of Magnetic Berezinskii-Kosterlitz-Thouless Physics}
\author{Mark Potts}
\affiliation{Max Planck Institute for the Physics of Complex Systems, N\"{o}thnitzer Str. 38, Dresden 01187, Germany}
\author{Shu Zhang}
\affiliation{Collective Dynamics and Quantum Transport Unit, Okinawa Institute of Science and Technology Graduate University, 1919-1 Tancha, Onna-son 904-0495, Japan}
\affiliation{Max Planck Institute for the Physics of Complex Systems, N\"{o}thnitzer Str. 38, Dresden 01187, Germany}
\maketitle

\section{Emergent maxwell equations}
 In this section we present a review of how the $(2+1)$d electromagnetic description of the XY model is obtained. We broadly follow the treatment found in Ref. \cite{Cote86}.
 
Starting from our model Hamiltonian:
\begin{equation}
    H=\frac{J_0}{2S^2}\sum_{i = x,y}  \int \text{d}^2\mathbf{r} \, (\mathbf{\nabla} S_i)^2 + \frac{1}{2\alpha}\int\text{d}^2\mathbf{r} \ S_z^2,
    \label{eq:supp_Hamiltonian}
\end{equation}
one can write down the equations of motion for the azimuthal angle variable $\phi$ and its conjugate momentum $S_z=\Pi_{\phi}\equiv \alpha \dot{\phi}$ as:
\begin{align}
\dot{\Pi}_{\phi} =& J_0 \nabla^2\phi , \\
\dot{\phi} =& \frac{1}{\alpha} \Pi_{\phi} .
\end{align}
We can rewrite the first of these as a wave equation:
\begin{equation}
\ddot{\phi} = \frac{J_0}{\alpha} \nabla^2\phi,
\end{equation}
describing the propagation of spin-wave excitations with bare speed $c_0=\sqrt{J_0/\alpha}$. 

Vortices are topological defects (homotopy defects) in the field $\phi$. To include their effects on the dynamics, one separates $\phi(\mathbf{r})$ into a non-compact field $\theta(\mathbf{r})$ associated with spin-waves, and a field $\psi(\mathbf{r})$ associated with the winding of $\phi$ about vortex cores. The field about a positive vortex centered at the origin is given by:
\begin{equation}
\psi(\mathbf{r}) = \arctan\left(\frac{y}{x}\right).
\end{equation}
If one inserts vortices with sign $n_i=\pm1$ at points $\mathbf{r}_i$, and then take the gradient of the resulting expression, one obtains:
\begin{align*}
\psi(\mathbf{r}) =& \sum_i n_i \arctan\left(\frac{y-y_i}{x-x_i}\right), \\
\mathbf{\nabla}\psi =& \sum_i n_i \nabla\arctan\left(\frac{y-y_i}{x-x_i}\right), \\
=& \sum_i n_i \frac{1}{|\mathbf{r}-\mathbf{r}_i|^2}\left\{(x-x_i)\hat{e}_y-(y-y_i)\hat{e}_x\right\}, \\
=& \sum_i n_i \hat{\mathbf{e}}_z\times \mathbf{\nabla}G(\mathbf{r}-\mathbf{r}_i).
\end{align*}
In the last line, the expression is rewritten in terms of the Green's function for the Laplacian in two dimensions, which satisfies:
\begin{equation}
\nabla^2G(\mathbf{r}-\mathbf{r}_i)=2\pi\delta(\mathbf{r}-\mathbf{r}_i).
\end{equation}
This Green's function is given by $\ln(|\mathbf{r}-\mathbf{r}_i|) +c$. If each vortex has a core radius of the order $a_0$, then $G(r) \sim \ln(r/a_0)$. As the curl of a gradient vanishes, one has that the vortex field satisfies $\mathbf{\nabla}\cdot\mathbf{\nabla}\psi = 0$. The vortex field is purely transverse, and trivially $\mathbf{\nabla}\theta$ -- the spin-wave part of $\mathbf{\nabla}\phi$ -- is purely longitudinal, allowing the Helmholtz decomposition:
\begin{equation}
\mathbf{\nabla}\phi=\mathbf{\nabla}\theta+\mathbf{\nabla}\psi .
\end{equation}
From stokes' theorem, one can show that the curl of $\mathbf{\nabla}\psi$ does not vanish due to the singularity at the center of each vortex, and is hence not a pure gradient. One instead finds that it satisfies:
\begin{equation}
\mathbf{\nabla}\times\mathbf{\nabla}\psi = \sum_i n_i 2\pi \delta(\mathbf{r}-\mathbf{r}_i)\hat{\mathbf{e}}_z 
\equiv 2\pi n(\mathbf{r}) \hat{\mathbf{e}}_z, \label{eq:vortex_grad_curl}
\end{equation}
where in the last line the vortex density $n(\mathbf{r})$ is defined. 

In taking the time derivative of the vortex field gradient, one must ensure the dynamics of the field singularities are properly accounted for. We define a vortex current $\mathbf{j}_v$ through the continuity equation:
\begin{equation}
\frac{\text{d}n}{\text{d}t} + \mathbf{\nabla}\cdot\mathbf{j}_v =0.
\end{equation}
Taking the time derivative of Eq. (\ref{eq:vortex_grad_curl}), and then integrating over a curve around a vortex, one finds:
\begin{align*}
\frac{\text{d}}{\text{d}t}(\mathbf{\nabla}\times\mathbf{\nabla}\psi)\cdot\hat{\mathbf{e}}_z =& -2\pi \mathbf{\nabla}\cdot\mathbf{j}_v, \\
\int_A \frac{\text{d}}{\text{d}t}(\mathbf{\nabla}\times\mathbf{\nabla}\psi)\cdot\mathbf{\text{d}A} =& -2\pi\int_A \mathbf{\nabla}\cdot\mathbf{j}_v \text{d}A, \\
\oint_{\partial A} \frac{\text{d}}{\text{d}t} \mathbf{\nabla}\psi\cdot\mathbf{\text{d}l} =& - 2\pi \oint_{\partial A} \mathbf{j}_v\cdot\mathbf{\text{d}s}.
\end{align*}
The line elements $\mathbf{\text{d}s}$ and $\mathbf{\text{d}l}$ are related by $\mathbf{\text{d}s}=\hat{\mathbf{e}}_z\times \mathbf{\text{d}l}$. Now equating the integrands of the last line to get the singular part of the time derivative, one finds that the contribution from the vortex cores is $-2\pi(\hat{\mathbf{e}}_z\times\mathbf{j}_v)$. The full time derivative is hence:
\begin{equation}
\frac{\text{d}}{\text{d}t}\mathbf{\nabla}\psi=\mathbf{\nabla}\left(\frac{\text{d}\psi}{\text{d}t}\right)-2\pi(\hat{\mathbf{e}}_z\times\mathbf{j}_v).
\end{equation}
The results so far obtained can be massaged into a form reminiscent of Maxwell's equations:
\begin{align}
\mathbf{\nabla}\cdot(\mathbf{\nabla}\phi\times\hat{\mathbf{e}}_z) =& 2\pi n(\mathbf{r}), \\ \mathbf{\nabla}\times(\mathbf{\nabla}\phi\times\hat{\mathbf{e}}_z) =& =-\frac{1}{J_0}\Pi_{\phi}\hat{\mathbf{e}}_z, \\ \mathbf{\nabla}\times(\Pi_{\phi}\hat{\mathbf{e}}_z) =& \alpha \frac{\text{d}}{\text{d}t}(\mathbf{\nabla}\phi\times\hat{\mathbf{e}}_z)+2\pi\alpha\mathbf{j}_v.
\end{align}
To make the analogy with electromagnetism clear, one can define an electric field proportional to $\mathbf{\nabla}\phi\times\hat{\mathbf{e}}_z$, and a magnetic field proportional to $\Pi_{\phi}\hat{\mathbf{e}}_z$ and thus recover the Maxwell equations in two dimensions. We define new quantities:
\begin{align}
\mathbf{E} =& \sqrt{2\pi J_0} \mathbf{\nabla}\phi\times\hat{\mathbf{e}}_z ,\\
\mathbf{B} =& \sqrt{\frac{2\pi}{\alpha}}\Pi_{\phi}\hat{\mathbf{e}}_z,\\
\rho(\mathbf{r}) =& \sqrt{2\pi J_0} \, n(\mathbf{r}), \\ \mathbf{j}(\mathbf{r}) =& \sqrt{2\pi J_0} \,\mathbf{j}_v(\mathbf{r}),
\end{align}
and using these definitions, the dynamical equations for spin-waves and vortices are mapped precisely to the Maxwell equations:
\begin{align}
\mathbf{\nabla}\cdot\mathbf{E} =& 2\pi \rho(\mathbf{r}), \\
\mathbf{\nabla} \cdot \mathbf{B} =&0, \\
\mathbf{\nabla}\times\mathbf{E} =& -\frac{1}{c_0} \frac{\partial\mathbf{B}}{\partial t}, \\ \mathbf{\nabla}\times\mathbf{B} =& \frac{1}{c_0} \frac{\partial \mathbf{E}}{\partial t} +\frac{2\pi}{c_0} \mathbf{j}(\mathbf{r}).
\end{align}

\section{Kinetic theory for bound and free vortices}
In this section, we provide more detail on our treatment of the kinetics of free and bound vortices used in our analysis, which primarily follows work presented in \cite{Ambegaokar79,Ambegaokar80,Huber82}. 

One can make use of the electromagnetic analogy to write down the equation of motion for a single vortex. This reads:
\begin{equation}
    m \ddot{\mathbf{r}} = 0 = q_i\sqrt{2\pi J_0} \mathbf{E} + \frac{q_i}{c_0} \sqrt{2\pi J_0} \dot{\mathbf{r}} \times \mathbf{B} -C \mathbf{v} -C' \hat{\mathbf{z}} \times \mathbf{v}.
\end{equation}
The forces on a vortex must balance as they possess no mass \cite{Cote86}. $q_i$ is the sign of the vortex (clockwise or anticlockwise winding/ positive or negative charge). The ratio of the magnitude of the electric and magnetic forces is the ratio of the spin-wave velocity $c_0$ to the vortex speed $v$ as expected for charged particles. 
$C$ and $C'$ parameterize drag forces acting on the vortex. \edit{Microscopically, these arise from Glibert spin damping. In the treatment in \cite{Huber82}, these drag coefficients are calculated directly from the canonical equations of motion of damped classical spins, where it is shown that $C' \approx 0$, and that $C \approx S\delta \log(L/a_0)$, with $S$ the magnitude of the spin density, $\delta$ the Gilbert spin-damping parameter, $L$ a macroscopic length scale, and $a_0$ the microscopic lattice spacing.}

\edit{Comparing the contributions of the Lorentz force and the linear drag force, one finds that they contribute in the ratio $ \sim|B|/S$. Our underlying assumption in this work is that spin dynamics are predominantly confined to the plane, and in this limit $B \propto S_z$. Hence, the ratio of the Lorentz and drag forces must also be small. Gilbert damping also in principle effects the dynamics of propagating spin waves, inducing a small frequency dependent spin-wave lifetime scaling with $\delta\omega$. Its contribution will not be affected by the unbinding of vortices at the BKT transition, and so provided this damping is sufficiently small that spin-waves are not overdamped below $T_c$, we can make the simplification of dropping this term without affecting the physics.}

\edit{With these considerations, one arrives naturally at the following drift velocity by dropping the Lorentz force contribution:}
\begin{equation}
    \mathbf{v}_d = q_i \frac{\sqrt{2\pi J_0}}{C} \mathbf{E}.
\end{equation}
Introducing a stochastic element to the motion via a Gaussian noise term $\eta(t)$, one arrives at the following Langevin equation:
\begin{equation}
    \frac{\text{d}\mathbf{r}_v}{\text{d}t} = q_i\frac{\sqrt{2\pi J_0}D}{k_BT} \ \mathbf{E} +\mathbf{\eta}(t), \label{eq:Master_Langevin}
\end{equation}
where $1/C$ has been re-parameterized as a mobility $D/k_BT=\nu$.
Equation \ref{eq:Master_Langevin} then serves as the basis for the kinetic theory of vortices used in our calculations, determining the linear response of bound pairs of vortices to changes in the surrounding spin configuration \cite{Ambegaokar79,Ambegaokar80,curtis2024probing}, and the equilibrium current density in the high temperature phase \cite{Ambegaokar80}.

Looking first at bound pairs, let us define the probability distribution $P(\mathbf{r},\mathbf{R},t)$ as the probability density for a pair of singly charged vortices of center-of-mass position $\mathbf{R}$ to have separation $\mathbf{r}=\mathbf{r}_p-\mathbf{r}_n$. The relevant Langevin equations for $\mathbf{r}$ and $\mathbf{R}$ are:
\begin{align}
    \frac{\text{d}\mathbf{r}}{\text{d}t} =& \frac{\sqrt{2\pi J_0}D}{k_BT} \ \left[\mathbf{E}(\mathbf{r}_1)+\mathbf{E}(\mathbf{r}_2)\right] +\mathbf{\eta}_r(t), \\
    \frac{\text{d}\mathbf{R}}{\text{d}t} =& \frac{\sqrt{2\pi J_0}D}{2 k_BT} \ \left[\mathbf{E}(\mathbf{r}_1)-\mathbf{E}(\mathbf{r}_2)\right] +\mathbf{\eta}_R(t),
\end{align}
with $\langle \eta_{r}^{\alpha}(t)\eta_r^{\beta}(t')\rangle = 4D \delta^{\alpha\beta}\delta(t-t')$, and $\langle \eta_{R}^{\alpha}(t)\eta_R^{\beta}(t')\rangle = D\delta^{\alpha\beta}\delta(t-t')$. Let us now also introduce a perturbing field $\delta \mathbf{E}(\mathbf{r},t)= i\mathbf{k}\delta V e^{\mathrm{i} \mathbf{k}\cdot\mathbf{r}}e^{-\mathrm{i} \omega t}$  to which we will calculate the linear response of the polarization (and hence the contribution to the dynamical dielectric constant). Including the coulomb interaction between the two vortices as well as this perturbation, the above Langevin equations become:
\begin{align}
    \frac{\text{d}\mathbf{r}}{\text{d}t} =& \frac{2\sqrt{2\pi J_0}D}{k_BT} \ \left[-\frac{\sqrt{2\pi J_0}\mathbf{r}}{r^2\epsilon(r)} +i\mathbf{k}\delta V(\mathbf{R},t)\cos\left(\frac{\mathbf{k}\cdot\mathbf{r}}{2}\right)\right] +\mathbf{\eta}_r(t), \\
    \frac{\text{d}\mathbf{R}}{\text{d}t} =& -\frac{\sqrt{2\pi J_0}D}{k_BT} \  \mathbf{k}\delta V(\mathbf{R},t)\sin\left(\frac{\mathbf{k}\cdot\mathbf{r}}{2}\right)+\mathbf{\eta}_R(t).
\end{align}
Here $\epsilon(r)$ is the scale dependent dielectric function derived from the BKT renormalization-group  equations \cite{Kosterlitz73,Kosterlitz74,Young78}. The corresponding Fokker-Planck equation for the probability distribution $P$ is then:
\begin{align}
    \frac{\partial P}{\partial t} =& \ \mathbf{\nabla}_{\mathbf{r}}\cdot\left\{\left(\frac{4\pi J_0 D\mathbf{r}}{k_BTr^2\epsilon(r)} - \frac{2D\sqrt{2\pi J_0}i\mathbf{k}\delta V}{k_BT}\cos\left(\frac{\mathbf{k}\cdot\mathbf{r}}{2}\right)\right)P\right\} \nonumber \\ &+\mathbf{\nabla}_{R}\cdot\left\{\left(\frac{\sqrt{2\pi J_0}D}{k_BT}\mathbf{k}\delta V\sin\left(\frac{\mathbf{k}\cdot\mathbf{r}}{2}\right)\right)P\right\} \nonumber \\ 
    &+ 2D\frac{\partial^2P}{\partial r^2}+\frac{D}{2}\frac{\partial^2P}{\partial R^2}.
\end{align}

\begin{figure*}
    \centering
    \includegraphics[scale=0.4]{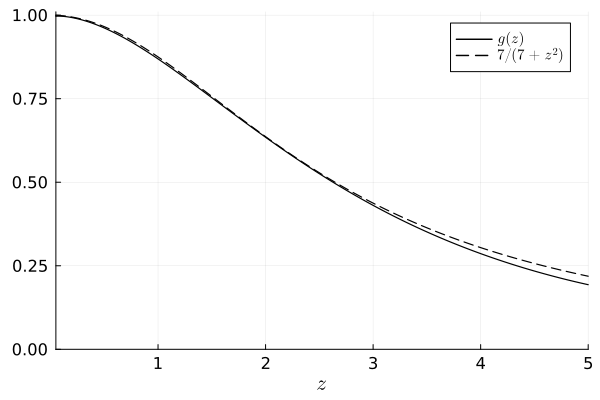}
    \caption{Comparison of the exact solution of Eq. (\ref{eq:Simple_g_ODE} for the boundary conditions $g(0)=1$, $\lim_{z\rightarrow \infty} g(z)=0$ with the function $7/(7+z^2)$ used to approximate $g(z)$ in Ref. \cite{Ambegaokar79,Ambegaokar80}. Whilst the differential equation for $g(z)$ we derive conflicts with that presented in Ref. \cite{Ambegaokar79}, the approximation for $g(z)$ adopted in that paper remains suitable despite the apparent error in that analysis.}
    \label{fig:New_g_plot}
\end{figure*}

We now separate out $P$ into the equilibrium distribution $P_0$, and the response $\delta P = \delta P(\mathbf{r}) e^{\mathrm{i} \mathbf{k} \cdot \mathbf{R}}e^{-\mathrm{i} \omega t}$. Keeping only terms linear in $\delta$, one obtains:
\begin{align}
    -\mathrm{i} \omega\delta P=& \mathbf{\nabla}_{r}\cdot\left\{\frac{4\pi J_0 D \mathbf{r}}{k_BT r^2\epsilon(r)}\delta P\right\} -i\frac{2D\sqrt{2\pi J_0}}{k_BT}k\delta V \cos\Theta\frac{\partial}{\partial r}\left\{\cos\left(\frac{\mathbf{k}\cdot\mathbf{r}}{2}\right)P_0\right\} \nonumber \\
    &+ \mathrm{i} k^2 \frac{\sqrt{2\pi J_0}D}{k_BT}\delta V\sin\left(\frac{\mathbf{k}\cdot\mathbf{r}}{2}\right)P_0 \nonumber\\
    &+2D\frac{\partial^2\delta P}{\partial r^2}-\frac{Dk^2}{2}\delta P.\label{eq:First_order_delta_P}
\end{align}
The angle $\Theta$ here is that between the perturbing field (or equivalently $\mathbf{k}$) and the relative position vector $\mathbf{r}$.
To further simplify this differential equation, one can make the following substitution for $\delta P$:
\begin{equation}
    \delta P = 2 \mathrm{i} \frac{\sqrt{2\pi J_0}}{k_BT}\sin\left(\frac{\mathbf{k}\cdot\mathbf{r}}{2}\right)\delta V P_0 g(r,\Theta,\omega,k) .\label{eq:Delta_P_ansatz}
\end{equation}
Inserting this into Eq. (\ref{eq:First_order_delta_P}), one can make use of the fact that $P_0$ satisfies:
\begin{equation}
    0= \frac{4\pi J_0 D}{k_B T}\mathbf{\nabla}_{\mathbf{r}}\cdot \left\{\frac{\mathbf{r}P_0}{r^2\epsilon(r)}\right\} +2D \frac{\partial^2 P_0}{\partial r^2}, \label{eq:Equilibrium_dist}
\end{equation}
which implies $\partial_r P_0 = -(P_0/k_BT) \partial_r U = -2\pi J_0 P_0/k_B T r \epsilon(r)$, to simplify the differential equation down to:
\begin{multline}
    r^2g'' +rg' \left\{2 \left(\frac{\mathbf{k}\cdot\mathbf{r}}{2}\right)\cot\left(\frac{\mathbf{k}\cdot\mathbf{r}}{2}\right) - \frac{2\pi J_0}{k_BT \epsilon(r)} \right\}  \\ +(1-g)\left\{\frac{k^2r^2}{4}(1+\cos^2\Theta)+\frac{2\pi J_0}{k_BT \epsilon(r)}\left(\frac{\mathbf{k}\cdot\mathbf{r}}{2}\right)\cot\left(\frac{\mathbf{k}\cdot\mathbf{r}}{2}\right)\right\} +\frac{\mathrm{i} \omega r^2}{2D}g=0.
\end{multline}
$g'$ and $g''$ indicate partial differentiation with respect to $r$. This equation is subject to the boundary conditions $g(0)=1$ and $\lim_{z\rightarrow\infty}g(z)=0$, and if one seeks only static solutions ($\omega=0$), then $g=1$ is the unique solution, which recovers the results of Ref. \cite{curtis2024probing}. If we take the limit $\mathbf{k} \rightarrow 0$, in principle we should recover the differential equation obtained in Ref. \cite{Ambegaokar79}. However we find instead the following slightly different equation:
\begin{equation}
    r^2g'' +rg'\left\{2-\frac{2\pi J_0}{k_BT \epsilon(r)}\right\} -g\left\{-\frac{\mathrm{i} \omega r^2}{2D} + \frac{2\pi J_0}{k_BT \epsilon(r)}\right\} +\frac{2\pi J_0}{k_BT \epsilon(r)}=0 . \label{eq:g_differntial_eq}
\end{equation}
The only difference between the equation arrived at in Ref. \cite{Ambegaokar79} and ours is that the `3' in the term proportional to $g'$ is replaced with a `2' in our equation. The method of analysis is identical here to as presented in Ref. \cite{Ambegaokar79}, and so we attribute the discrepancy to an error made in the original paper. As in that paper, we can further approximate Eq. (\ref{eq:g_differntial_eq}) by replacing the combination $2\pi J_0/k_BT \epsilon(r)$ by its bulk critical value of $4$, and then defining $z^2=-\mathrm{i}\omega r^2/2D$, giving:
\begin{equation}
    z^2g''-2zg'-(z^2+4)g+4=0. \label{eq:Simple_g_ODE}
\end{equation}
The dynamical dielectric constant $\epsilon(\omega,\mathbf{k})$ is defined in terms of the change in the distribution function $\delta P$ as \cite{Ambegaokar79}:
\begin{equation}
    \epsilon(\omega,k) = 1 + 4\pi \sqrt{2\pi J_0} \int \text{d}^2\mathbf{r} \frac{\mathbf{r}\cdot \mathbf{k}}{2 \mathrm{i} k^2} \frac{\delta P}{\delta V}.
\end{equation}
Inserting Eq. (\ref{eq:Delta_P_ansatz}) into the above equation, and using the self-consistency equation for the scale-dependent dielectric constant $\epsilon(r)$ \cite{Kosterlitz73,Kosterlitz74,Young78}, one obtains:
\begin{equation}
    \epsilon(\omega,k) = 1 +\int_{a_0}^{\infty} \text{d}r\frac{\text{d} \epsilon(r)}{\text{d} r} \int_0^{2\pi}\text{d}\Theta\left[\frac{2\cos\Theta}{\pi k r}\sin\left(\frac{kr\cos\Theta}{2}\right)\right]g(r,\Theta,\omega,k).
\end{equation}
In the main text, we make the same approximation as made in Ref. \cite{curtis2024probing}, and evaluate $g$ in the $\mathbf{k} \rightarrow 0$ limit, wherein it is independent of both $k$ and $\theta$, and the above expression simplifies to:
\begin{equation}
    \epsilon(\omega,k) = 1 +\int_{a_0}^{\infty} \text{d}r\frac{\text{d} \epsilon(r)}{\text{d} r} \left[\frac{4}{kr}\mathcal{J}_1\left(\frac{kr}{2}\right)\right]g(r,\omega) \approx 1 +\int_{a_0}^{\infty} \text{d}r\frac{\text{d} \epsilon(r)}{\text{d} r} e^{-k^2r^2/32}g(r,\omega).
\end{equation}
Here $\mathcal{J}_1(z)$ is a Bessel function of the first kind. To arrive at the equation in the main text, all that remains is to solve Eq. (\ref{eq:Simple_g_ODE}) for $g(r,\omega)$. A particular solution to this equation that satisfies the required boundary conditions is the following:
\begin{equation}
    g(z) = \frac{1}{2z}\left\{\text{Shi}(z)\left((z^2+3)\cosh(z)-3z\sinh(z)\right)-\text{Chi}(z)\left((z^2+3)\sinh(z)-3z\cosh(z)\right)-z\right\},
\end{equation}
where $\text{Shi}(z)$ and $\text{Chi}(z)$ are the hyperbolic sine and cosine integral functions respectively. Comparing this expression for $g(z)$ with the approximate form suggested in Ref. \cite{Ambegaokar79} of $g(z)=c/(c+z^2)$ for some constant $c$ -- as shown in Fig. \ref{fig:New_g_plot}-- we find that this corrected $g(z)$ is still well approximated by $g(z) \sim 7/(7+z^2)=(14D/r^2)/(14D/r^2-\mathrm{i} \omega)$.

We now turn our attention to the situation above $T_c$, and to the free vortex current at equilibrium. Following \cite{Ambegaokar80} we use Eq. (\ref{eq:Master_Langevin}) and define the free vortex charge density $\rho_\text{free}=\sqrt{2\pi J_0}(n_{+}-n_{-})$. One obtains the following Fokker-Planck equation for the free charge density:
\begin{equation}
    \frac{\partial \rho_{\text{free}}}{\partial t} =-\frac{2\pi J_0 D}{k_BT} \mathbf{\nabla}_r \cdot \left[ (n_++n_-)\mathbf{E} \right] +D \nabla^2\rho_{\text{free}}.
\end{equation}
Approximating $n_++n_-$ with a constant total vortex density $n_f$, and defining $\sigma=2\pi J_0 D n_f/k_BT$, one can identify the following conserved current density:
\begin{equation}
    \mathbf{j}_\text{free} =\sigma \mathbf{E} - D\mathbf{\nabla}\rho_\text{free}.
\end{equation}
Moving to Fourier space, and using $\partial_t \rho_{\text{free}} = - \mathbf{\nabla}\cdot\mathbf{j}$ to eliminate the charge density, one then obtains the expression for the current density used in the main text above $T_c$, in terms of longitudinal and transverse electric fields:
\begin{equation}
    \mathbf{j}_{\text{free}}=\sigma\frac{\mathrm{i} \omega}{\mathrm{i} \omega-Dk^2}\mathbf{E}_L+\sigma\mathbf{E}_T.
\end{equation}

In addition to the presence of a free current, and it's effect on the linear response of the system, the expression for the dynamical dielectric constant must also be modified. Above $T_c$, the spacing between unbound vortices is of the order of the correlation length $\xi_+$, and it is only below this length scale that it is sensible to discuss bound pairs. Thus the integration limits on the dynamical dielectric constant must be altered to read \cite{Ambegaokar80}:

\begin{equation}
    \epsilon_{b}(\omega,k)= 1 + \int_{a_0}^{\xi_+} \text{d} r \frac{\text{d}\epsilon(r)}{\text{d} r} e^{-k^2r^2/32}g(r,\omega). \label{eq:Above_Tc_dielectric_constant}
\end{equation}

As $\xi_+ \rightarrow a_0$, $\epsilon_b(\omega,k)$ becomes approximately independent of $\omega$ and $k$, and saturates at $\epsilon_c$, the critical bulk value.

\section{Linear response of spin waves and vortices}

To calculate the spin-density correlation functions presented in the main text, we follow Ref.~\cite{Cote86} and Ref.~\cite{Lifshitz80} and obtain first the relevant linear response functions, and relate them back to the desired correlation functions using the fluctuation-dissipation theorem. 

The most convenient setting for calculating these response functions is in the electromagnetic picture, where derivatives of the order parameter $\phi$ become electric and magnetic fields:
\begin{align}
    \mathbf{E}_L=\sqrt{2 \pi J_0} \ \mathbf{\nabla} \psi \times \hat{\mathbf{e}}_z, \label{eq:Longitudinal E field}\\
    \mathbf{E}_T=\sqrt{2 \pi J_0} \ \mathbf{\nabla} \theta \times \hat{\mathbf{e}}_z, \\
    \mathbf{B} = \sqrt{2 \pi \alpha} \ \dot{\phi} \  \hat{\mathbf{e}}_z. \label{eq:B field}
\end{align}

The correlations of both the electric and magnetic fields can both be derived from those of the vector potential $\mathcal{A}$, which we calculate in the Weyl (zero potential) gauge. As discussed in Ref.~\cite{Lifshitz80}, correlations of the vector potential can be obtained from the imaginary part of the retarded Green's function for the vector potential $\mathcal{G}^{R}_{\mu,\nu}(\omega,\mathbf{r})$, which relates a current source to the macroscopic vector potential
\begin{equation}
    A_{\mu}(\omega,\mathbf{r}) = -\frac{1}{\hbar c} \int  \mathcal{G}^{R}_{\mu,\nu}(\omega,\mathbf{r}-\mathbf{r}') \bar{j_{\nu}}(\omega,\mathbf{r}') \text{d}^2\mathbf{r}'. \label{eq:Linear_response}
\end{equation}
The current source $\bar{\mathbf{j}}$ should here be distinguished from the equilibrium current $\mathbf{j}_\text{eq}$ due to the motion of free vortices; $\bar{\mathbf{j}}$ describes some fluctuation about equilibrium.

In Fourier-transformed variables, the Maxwell equations in a dielectric relating the vector potential to the total current read:
\begin{align}
    \left[k^2 - \frac{\omega^2}{c_0^2} \epsilon(\omega,k)\right] A_{T}(\omega,k) =& \frac{2\pi}{c_0} \left(j_{\text{eq};T} + \bar{j_{T}} \right), \\
    - \frac{\omega^2}{c_0^2} \epsilon(\omega,k) A_{L}(\omega,k) =& \frac{2\pi}{c_0} \left(j_{\text{eq};L} + \bar{j_{L}} \right).
\end{align}
Below the critical temperature, the approximation is made to account for the effects of bound vortices through the dynamical dielectric function alone, and consider only spin-wave fluctuations \cite{Cote86}. This is achieved by taking the spin-wave field $\theta_0$ that obeys the vortex-free Maxwell equations with a dynamical dielectric constant. $\theta_0$ accounts for all the dynamics in $\phi$ below $T_c$, and is related to the emergent electric field by
\begin{equation}
    \epsilon(\omega,k)\mathbf{E}_T  = \sqrt{2\pi J_0} \nabla \theta_0 \times \hat{\mathbf{e}}_z .
\end{equation}
Thus in the BKT phase, in the absence of free vortices, the longitudinal component of the Green's function can be discarded, and with no equilibrium currents, one obtains:
\begin{equation}
    \mathcal{G}^R_{T}(\omega,k) = \frac{2\pi \hbar} {(\omega^2/c_0^2)\epsilon(\omega,k)-k^2}.
\end{equation}
Note that factors of $\hbar$ here arise due to the use of the Kubo formula in arriving at Eq.~(\ref{eq:Linear_response}). We are interested in the classical limit, and will take $\hbar \rightarrow 0$ when appropriate. Using the fluctuation-dissipation theorem for bosonic particles, the fluctuations in the transverse vector potential are then:
\begin{equation}
    \mathcal{C}_{T}(\omega,k)= - \coth\left(\frac{\hbar \omega}{2 k_BT}\right) \text{Im} \{ \mathcal{G}^R_{T}(\omega,k)\}.
\end{equation}
Focusing on the regime $\hbar \omega \ll k_B T$, one returns to the classical limit. The transverse vector potential correlations can then be related to the correlation functions for the field $\phi$ (approximated by those of $\theta_0$) and the out-of-plane spin density via:
\begin{align}
    \mathcal{C}_{\theta_0}(\omega,k) &\approx \frac{\omega^2\epsilon_{\text{Re}}(\omega,k)^2}{2\pi J_0 k^2 c_0^2}\mathcal{C}_{T}(\omega,k), \label{eq:theta_correlator}\\
    \mathcal{C}_{z}(\omega,k) &= \frac{J_0 k^2}{2\pi c_0^2}\mathcal{C}_{T}(\omega,k),
\end{align}
by making use of the Fourier transforms of the definitions  (\ref{eq:Longitudinal E field}-\ref{eq:B field}). We have used the approximation that $\epsilon \approx \epsilon_\text{Re}$  in Eq.~(\ref{eq:theta_correlator}), as $\epsilon_{\text{Im}} \ll 1$ is found by evaluation of the renormalization-group equations for a bare vortex chemical potential $\mu_0$ larger than the spin-stiffness $J$. 
Explicit calculation of the vortex-antivortex creation energy shows that this is the case~\cite{Kosterlitz73}.

The in-plane spin-density correlations are given in terms of the $\phi$ correlations by:
\begin{equation}
\mathcal{C}_{\perp}(t,\mathbf{r})
    = S^2e^{\mathcal{C}_{\phi}(t,\mathbf{r})-\mathcal{C}_{\phi}(0,\mathbf{0})} .
\end{equation}
To evaluate this, the approximation $\epsilon_\text{Im} \approx 0$ is employed, and one finds below $T_c$:

\begin{equation}
\mathcal{C}_{\phi}(t,\mathbf{r})\approx C_{\theta_0}(t,\mathbf{r}) \approx \frac{k_BT}{2\pi} \int_0^{\infty} \text{d}k \frac{\epsilon_{\text{Re}}(\omega,k)}{J_0 k}\mathcal{J}_0(kr)\cos(\omega t)
\end{equation}
Here $\mathcal{J}_0(z)$ is a bessel function of the first kind.

Looking now above $T_c$, we can make use of the manipulations discussed in the previous section to include the effects of free vortices in the high spin-damping limit. We have
\begin{equation}
\mathbf{j}_\text{eq}(\omega,k)=\sigma\frac{\mathrm{i} \omega}{\mathrm{i} \omega-Dk^2}\mathbf{E}_L+\sigma\mathbf{E}_T.
\end{equation}
Following the same manipulations as for below $T_c$, the transverse fluctuations are, in full:
\begin{equation}
    \mathcal{C}_{T}(\omega,k) =  \frac{4\pi k_BT  c^2 \ (2\pi \sigma + \omega\epsilon_{\text{Im}}(\omega,k)) }{\left[\omega^2\epsilon_{\text{Re}}(\omega,k)-k^2c_0^2\right]^2 + \left[2\pi\sigma\omega+\omega^2\epsilon_{\text{Im}}(\omega,k)\right]^2}.
\end{equation}
The dynamical dielectric function includes only the effects of bound pairs, and is approximately constant $(\sim \epsilon_c)$ and sufficiently far above $T_c$. The longitudinal correlations are:
\begin{equation}
    \mathcal{C}_{L}(\omega,k)  =\frac{ \left(4\pi k_BT c^2/\omega^3\right) \left[2\pi \sigma \omega + \epsilon_{\text{Im}}(\omega,k)(\omega^2+(Dk^2)^2)\right] [(Dk^2)^2+\omega^2]}{\left\{\epsilon_c[(Dk^2)^2+\omega^2]+2\pi\sigma Dk^2\right\}^2+\left\{2\pi \sigma \omega + \epsilon_{\text{Im}}(\omega,k)[\omega^2+(Dk^2)^2]\right\}^2}.
\end{equation}
In practice, we make use of the approximation $\epsilon \sim \epsilon_c$ when evaluating the longitudinal correlations.

The mapping from the longitudinal vector potential correlations and the $\psi$ field correlations is not straight-forward. This is touched on in Ref.~\cite{Cote86}: whilst $\nabla \psi$ is a well behaved function, $\psi$ winds around each vortex, and is singular at each vortex core (allowing $\psi$ to violate Stokes' theorem). The $\psi$ field of a single vortex only has angular dependence, and must either contain branch cuts or be multi-valued. The Fourier-transform of this function is thus particularly ill-defined, and so the step $C_{\psi} = (\omega^2/2\pi J_0 k^2 c^2) C_{L}$ is not valid. However, as our system always contains a net zero vorticity, at large distances $\psi$ must be a constant function, avoiding singularities in taking the Fourier transform at large scales. This does not solve the problem of branch-cuts or multiple values about each vortex however, and so we can only use $C_{L}$ to approximate $C_{\psi}$ up to a cut off momentum corresponding to length scales similar to the vortex spacing $\sim \xi_+$. This in turn implies a temperature scale above which our analysis is accurate for a given range of momenta, set by requiring $1/\xi_+ \gg 1/d$, which ensures no contribution to the relaxation rate from wave-numbers outside the valid range.

Above $T_c$, we have independent contributions from spin-waves and vortices (there are no off-diagonal components of the Green's function in the strong damping limit, and so vortices and spin-waves are decoupled). $C_{\theta_0}(t,\mathbf{r})$ and $C_{\psi}(t,\mathbf{r})$ are approximated by:
\begin{align}
C_{\theta_0}(t,\mathbf{r}) \approx& \frac{k_BT \epsilon_c}{2\pi J_0} \int_0^{\infty} \text{d}k \frac{1}{k}\mathcal{J}_0(kr) e^{-\pi\sigma t/\epsilon_c} \left\{\cos\left[\frac{\Delta(k)t}{2}\right] - \frac{2\pi\sigma}{\epsilon_c \Delta(k)}\sin\left[\frac{\Delta(k)t}{2}\right]\right\}, \\
C_{\psi}(t,\mathbf{r}) \approx& \frac{k_BT}{2\pi J_0} \int_0^{\infty} \text{d}k \frac{2\pi \sigma}{k(2\pi\sigma/\epsilon_c + Dk^2)} \mathcal{J}_0(kr) e^{-(2\pi\sigma/\epsilon_c + Dk^2)t}, \\
\Delta(k) =& \sqrt{4k^2c_0^2/\epsilon_c - (2\pi\sigma/\epsilon_c)^2}.
\end{align}

The removal of propagating spin-wave modes can be seen at the level of these correlation functions above $T_c$ through the behavior of the function $\Delta(k)$. For $kc \gg \sigma$, this function is real, and so the integrand is oscillatory; in the limit $\sigma \rightarrow 0$ we recover the result for below $T_c$, albeit with a constant dielectric constant. For large $\sigma$ however, $\Delta(k)$ is imaginary, and thus the integrand decays exponentially in time. We note also that in this limit of large $\sigma$ (large vortex density), $C_{\theta_0}(t,\mathbf{r})$ and $C_{\psi}(t,\mathbf{r})$ are identical.

\begin{figure*}[b]
    \centering
    \includegraphics[scale=0.16]{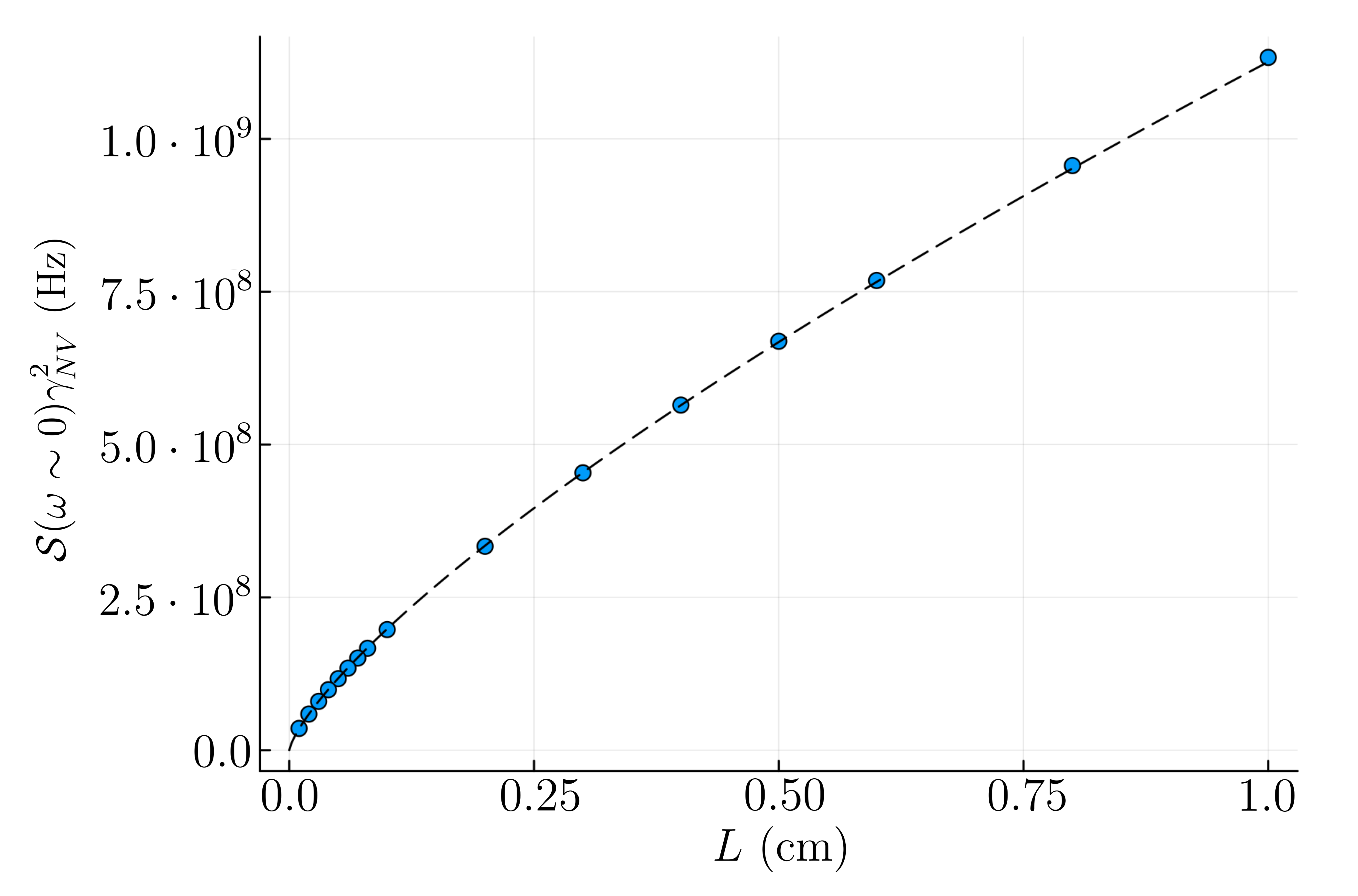}
    \caption{Scaling of near zero value of $\mathcal{S}(\omega)$ calculated for increasing values of the linear system size $L$. Finite size effects are incorporated as a distance/time scale introducing exponential decay of correlations at late times and large distances. The resulting scaling with $L$ is fitted to a power-law $\mathcal{S}(\omega \sim 0) = AL^B$, with $B=0.754$, in close agreement with a rough calculation which predicts an exponent of $1-\eta \approx 0.761$ at the chosen temperature of $T=0.97T_c$. }
    \label{fig:Finite_scaling_plot}
\end{figure*}

\section{Na\"ive finite size scaling in low temperature phase}

In the low-temperature phase, as discussed in the main text, $\mathcal{S}(\omega)$ inherits algebraic low-frequency behavior from the correlations of the order parameter. In practice, these algebraic correlations will be sensitive to finite-size effects and other perturbations which introduce additional length scales into the system.

For a system of linear size $L$, the maximum coherence time is of the order $L/c_0$. The frequency dependence of $\mathcal{S}(\omega)$ at frequencies below the spin-wave maximum at $\omega \sim c_0/d$ is determined approximately by the integral:
\begin{equation}
    \int \text{d}t \cos(\omega t) (c_0 t)^{-\eta}e^{-c_0t/L},
\end{equation}
hence $\mathcal{S}(\omega \sim 0) \propto L^{1-\eta}$. Performing the integrals contributing to $\mathcal{S}(\omega)$ more precisely, the low frequency scaling presented in Fig. \ref{fig:Finite_scaling_plot} is obtained, which is fitted to a power law with exponent $0.754 \approx 1-\eta$ for the chosen temperature.

\bibliography{SupBibliography.bib}